# Impacts of Sulfate Injection Geoengineering on Particulate Matter with Diameter less than 2.5 µm

Lili Xia[1], Alan Robock[1], and Simone Tilmes[2]

[1]*Department of Environmental Sciences, Rutgers University, New Brunswick, NJ*
[2]*Atmospheric Chemistry Observations and Modeling Laboratory, National Center of Atmospheric Research, Boulder, CO*
*Correspondence to: Lili Xia (lxia@envsci.rutgers.edu)*

## Abstract

Particulate matter with aerodynamic diameter less than 2.5 μm (PM2.5) is of great concern for human health. Here, for the first time, we examine the impact of sulfate aerosol geoengineering on PM2.5, using the output from the Geoengineering Large Ensemble (GLENS) project. GLENS is an ensemble of climate model simulations injecting $SO_2$ into the stratosphere to balance RCP8.5 forcing using the Community Earth System Model, version 1. GLENS geoengineering reduces global averaged surface PM2.5 mass concentrations compared with RCP8.5 and also changes PM2.5 composition with more percentages of organic carbon and sulfate. The total reduction of PM2.5 is a result of less dust and sea salt concentrations. Dust emission is declined under GLENS geoengineering because of increased soil moisture and leaf area index over desert regions, and less emission of sea salt is due to slower wind speeds when geoengineering is applied. Excluding dust and sea salt, there is more global averaged PM2.5 under GLENS geoengineering relative to RCP8.5, predominantly due to more aerosol phase secondary organic aerosol (SOA). Since gas precursors of SOA are prescribed in the simulations, a cooler environment with geoengineering tends to transfer more gas phase SOA to aerosol phase. Changes in PM2.5 concentration and composition with applied geoengineering may have potential human health impact. Another new finding of this study is that the large amount of injected $SO_2$ does not increase surface sulfate aerosol as a component of PM2.5, as the majority of sulfate aerosol reaching the boundary layer is in the coarse mode, which represents a very small fraction of the PM2.5. But the difference of deposition spatial distribution between geoengineering and RCP8.5 may have potential impacts on ecosystem.

## I. Introduction

PM2.5 is one of the most dangerous air pollutants for public health (Ebi and McGregor, 2008; WHO, 2013). It has a wide variety of anthropogenic and natural sources and can be emitted directly or formed indirectly from gases in the air as the secondary aerosol. With less anthropogenic emissions under future scenarios, global averaged PM2.5 concentration tends to show a decreasing trend. Moreover, PM2.5 concentration is also highly influenced by climate changes (e.g., Fang et al., 2011; Xu and Lamarque, 2018; Jacob and Winner, 2009; Lam et al., 2011; Hedegaard et al., 2013; Von Schneidemesser et al., 2015). There have been several model studies on future climate change impacts on PM2.5, but the results are either not consistent in magnitude or even contradictory to each other (Jacob and Winner, 2009). For example, without considering emission changes, some show that global warming may slightly reduce PM2.5 air pollution as a result of increased precipitation (Racherla and Adams, 2006; Colette et al., 2013), whereas others show that, with more rainfall under global warming, changes in precipitation frequency and intensity may increase reginal PM2.5 concentrations (Pye et al., 2009; Jacob and Winner, 2009; Xu et al., 2018).

Solar radiation management is one of the mostly discussed potential schemes to reduce anthropogenic warming and gain time for society to ramp up more effective mitigation pathways. We refer to the scheme studied here as sulfate aerosol geoengineering (SAG) - $SO_2$ is emitted as a gas, and the aerosol forms in situ from the gas. On one hand, SAG can affect surface PM2.5 concentrations directly due to the fact that sulfate aerosols themselves settle from the stratosphere into the troposphere, where they are either removed through precipitation (wet removal) or deposited at the surface (dry deposition). On the other hand, SAG induced climate changes can influence the concentrations of all the aerosols. There has been only one study on how SAG affects sulfate-aerosol-related PM2.5 (Eastham et al., 2018) using an offline global chemical-transport model. They found that to offset a 1°C anthropogenic warming, 1 TgS will be injected every year. With this extra sulfate aerosol and reduced temperature, surface PM2.5 concentration will increase and result in greater mortality. However, this study didn't consider the contribution of other components of PM2.5 including fine dust, fine sea salt, secondary organic matter, black carbon and primary organic matter. There are also two studies focusing on the deposition rate change of sulfate aerosol in total (Kravitz et al., 2009; Visioni et al., 2018), which showed that SAG would enhance the sulfate deposition, corresponding to the amount of injection, and would change the regional deposition patterns.

This is the first comprehensive study on how SAG would affect surface PM2.5 concentration and its composition by using output from a fully coupled earth system model with a modal aerosol module. We also investigate the regional pattern changes in PM2.5 concentration under SAG and try to understand the potential human health impacts.

## II. Methods

We use model output from the stratospheric aerosol Geoengineering Large Ensemble Project (GLENS; Tilmes et al., 2018). This project used Representative Concentration Pathway 8.5 (RCP8.5; Meinshausen et al., 2011) as the reference case from 2010 to 2099, and SAG starts in 2020. The sulfate injection locations are at the latitude bands of 30°N, 30°S, 15°N and 15°S and the injection height is ~5 km above the tropopause. The injection amount at each of the four locations is adjusted yearly by a feedback-control algorithm (Kravitz et al., 2017; MacMartin et al., 2017) to maintain (1) the global mean surface temperature, (2) the interhemispheric temperature gradient and (3) the equator-to-pole temperature gradient to be the same as in year 2020. The total injected aerosol amount increases from 0 Tg $SO_2$ $a^{-1}$ in 2020 to 53 Tg $SO_2$ $a^{-1}$ in 2100 (Tilmes et al., 2018). GLENS includes 20 ensemble members of SAG simulations (2020-2099), 20 ensemble members of RCP8.5 from 2010 to 2030, and 3 ensemble members of RCP8.5 from 2020 to at least 2097. In this study, 20 ensemble members of RCP8.5 between 2010 and 2019 serve as the control.

All simulations are performed by the Community Earth System Model, version 1, using the Whole Atmosphere Community Climate Model as its atmospheric component (CESM1(WACCM)) with a horizontal resolution in latitude and longitude of 0.9°×1.25°, respectively, and 70 vertical layers (Mills et al., 2017). The model includes a modal aerosol module with three lognormal modes (MAM3; Liu et al., 2012), which is capable of simulating the aerosol size distribution and both internal and external mixing between aerosol components, and of treating complicated aerosol processes. The model has comprehensive middle atmosphere chemistry with a simplified chemistry scheme in the troposphere, which supports the formation of aerosols. Sulfate aerosol in the troposphere is formed by (1) $SO_2$ oxidation with $H_2O_2$ and $O_3$; (2) $SO_2$ oxidation with OH to form gas phase $H_2SO_4$, and then gas phase $H_2SO_4$ is irreversibly condensed on cloud droplets with $NH_3$ to form $(NH_4)_2SO_4$ (Neale et al., 2010). WACCM is coupled with the Community Land Model version 4.5 (CLM4.5) with biogeochemistry turned on.

MAM3 outputs three modes of aerosols including sulfate, secondary organic aerosol (SOA), black carbon (BC), primary organic matter (POM), dust and sea salt (Table 1). Nitrogen-related aerosol is not a direct output variable in MAM3, but sulfate is partially neutralized by ammonium in the form of $NH_4HSO_4$, so that ammonium is effectively prescribed (Liu et al., 2012). In this study, we analyze mass concentrations of PM2.5 components, and in the following text, we use "concentration" to indicate mass concentration. We calculate the contribution of each component to total PM2.5 concentration with the following equation (Seinfeld and Pandis, 2016) since PM2.5 has not been explicitly calculated in this model version:

$$\text{PM2.5} = \text{Aitken} + \text{Accumulation} + 0.1378 \times \text{Coarse}$$

## III. Results

SAG in GLENS successfully achieves the temperature objectives (Figure 1a) but with decreasing global averaged precipitation as the simulation progresses (Figure 1b), as shown in many previous studies (e.g., Jones et al., 2013). Precipitation over land is on average almost unchanged (Fasullo et al., 2018), while some regional changes over land are still large (Simpson et al., 2019 submitted). Global average surface PM2.5 shows a downward trend in both RCP8.5 and GLENS as a result of emission reductions from both anthropogenic and natural sources which are the same in both experiments (Figures 1c and 1d). Compared with RCP8.5, SAG results in less global averaged PM2.5 (Figure 1c), but it has more PM2.5 when dust and sea salt are excluded (Figure 1d). During the period of 2080-2089, RCP8.5 and SAG show similar spatial patterns of PM2.5 concentration changes compared with the control – strong reduction in arid regions and their surrounding areas, including Australia, Middle East, Central Asia, East Asia and Sahara (Figures 2a and 2b). When dust and sea salt are not included, the reductions of PM2.5 in RCP8.5 and SAG are over all continents, particular in the Northern Hemisphere, as a result of less anthropogenic emissions from high population regions (Figures 2d and 2e). Compared with RCP8.5 (2080-2089), PM2.5 reduction in SAG is stronger over arid regions with the largest reduction over Central Australia, Sahara and Central Asia, and some positive impacts in spotted areas of India and Middle East (Figure 2c). On the other hand, without the dust and sea salt influence, SAG tends to increase surface PM2.5 concentration relative to RCP8.5, mainly over the Tropics (Figure 2f). Although the absolute changes of PM2.5 concentration between SAG geoengineering and RCP8.5 are stronger over low and middle latitudes, the highest percentage changes are over remote regions, such as Greenland and Antarctic, with values larger than 100%. This increase of PM2.5 surface concentration as well as deposition in clean regions would potentially have significant impacts on local ecosystems.

In this study, PM2.5 is calculated from six components (Table 1). The largest mass concentration contributor is dust, which makes up 49% of the total PM2.5 concentration; second and third are sea salt (33%) and sulfate aerosol (9%); primary organic matter and secondary organic aerosol are 5% and 3% of total PM2.5, respectively; and black carbon is less than 1% (Figure 3a). All the above percentages are based on the averaged ensemble members of the control climatology (RCP8.5, 2010-2019). The percentages of components would change as a function of time and scenarios. Dust and sea salt are a smaller fraction of the total PM2.5 under SAG compared with RCP8.5 as a result of less absolute mass concentrations, while other components would all have higher percentages (Figure 3b). The percentage changes of PM2.5 components under SAG indicate potential health impacts as individual aerosols in PM2.5 have different influences on humans (see more discussion in section IV).

Figure 4 shows the concentrations of PM2.5 components and their wet and dry depositions in the three modes for the control RCP8.5 (2010-2019), RCP8.5 (2080-2089), and SAG (2080-2089). Aerosol wet deposition includes in-cloud scavenging which is removal of aerosols acting as cloud condensation nuclei and below-cloud scavenging which removes aerosols below cloud by precipitation (Rasch et al., 2000). The strength of wet deposition is determined by the precipitation rate, aerosol solubility, a scavenging coefficient, and a tuning factor in the model (Easter et al., 2004). Aerosol dry deposition depends on gravitational settling and cloud-borne aerosol sedimentation.

***Sulfate Aerosols***

Under RCP8.5, anthropogenic emission of $SO_2$ reduces from 100 Tg $SO_2$ $a^{-1}$ in 2020 to 25 Tg $SO_2$ $a^{-1}$ in 2100 (van Vuuren et al., 2011; Smith et al., 2001 and 2004). While in the GLENS project, additional $SO_2$ is injected to counteract the anthropogenic radiative forcing of RCP8.5 with amount of 53 Tg $SO_2$ $a^{-1}$ in the end of $21^{st}$ century (Tilmes et al., 2018). The sulfur dioxide interacts with other atmospheric components and forms sulfate aerosols through chemical and physical processes. Stratospheric sulfate aerosols have an *e*-folding lifetime of 1-2 years (Pitari et al., 2016; Mills et al., 2017), and the fallout impacts the surface sulfate aerosol concentration, as well as acid rain and snow.

The sulfate component of PM2.5 shows small increases under SAG compared with RCP8.5 of ~0.01 $\mu g/m^3$ based on our calculations (Figure 4). This increase is from coarse mode only, and sulfate aerosols in the Aitken and accumulation modes decrease under SAG geoengineering relative to RCP8.5 by 0.02 $\mu g/m^3$, primarily due to decreases in the accumulation mode. Sulfate aerosol surface emissions in Aitken and accumulation modes follow the same prescribed emission files under both SAG geoengineering and RCP8.5 – sulfate aerosols in the Aitken mode are from domestic and transportation sources and sulfate aerosols in the accumulation mode are from agriculture, waste, shipping, energy, industry, forest fire and grass fires (Neale et al., 2010). The model used in this study only includes simple gas-phase chemistry in the troposphere. Gas phase sulfate forms first, and then irreversibly transfers to aerosol phase sulfate. The new sulfate aerosol particles are added to the Aitken mode and move to accumulation mode by condensation and coagulation (Neale et al., 2010; Liu et al., 2012). Therefore, the differences in fine mode sulfate aerosol concentrations between SAG geoengineering and RCP8.5 are from the gas-phase chemistry, gas-aerosol exchange and the wet and dry depositions. Under SAG, the injected $SO_2$ turns into sulfate aerosol, and with higher concentration, it is easier to grow into larger particles through nucleation, accumulation and coagulation compared with RCP8.5.

Changes in wet deposition of sulfate aerosol, particular wet deposition below-cloud (the removal of interstitial aerosol particles by precipitation particles through impaction and Brownian diffusion) is consistent with precipitation changes. Figure 5 shows that reduced below-cloud wet scavenging in SAG matches the precipitation reduction pattern. In RCP8.5, the Intertropical Convergence Zone (ITCZ) shifts northward since the Northern Hemisphere has a stronger warming than Southern Hemisphere, and in SAG, the ITCZ is approximately in the same position as in the control (Cheng et al., 2019 submitted). Therefore, when comparing SAG to RCP8.5 during 2080-2089, there is more precipitation south of the Equator, while over and north of the Equator, there is a strong precipitation reduction (Figure 5a). The wet deposition below cloud reduces with the same pattern as precipitation reduction (Figure 5b).

Dry deposition depends on particle size, density of the particle, surface conditions, and meteorological conditions (Zhang et al., 2001). Since injected $SO_2$ forms more fine particles which tend to accumulate and coagulate into large particles, dry deposition of fine sulfate particles (Aitken and accumulation modes) under SAG is less than RCP8.5 (Figure 4), while dry deposition of coarse sulfate aerosols is more than RCP8.5, which dominates the dry deposition of sulfate aerosols in the total of the three modes (Figure 6d).

The total concentrations of sulfate aerosol at the surface (the sum of three modes) are similar in SAG and the control, since the amount of injected sulfate is equivalent to the reduction of anthropogenic sulfate emission in RCP8.5. However, the size distributions of sulfate aerosol in SAG and the control have changed significantly – in the control there are more sulfate aerosols in Aitken and accumulation modes, while under SAG there are more coarse mode sulfate aerosols. Therefore, even with the same surface sulfate concentration, finer sulfate particles for current conditions may have higher health risks. In addition, the global distribution of sulfate deposition is different in SAG and the control (Figure 6e). There is more sulfate deposition over remote clean regions in SAG than the control, including the Southern Ocean, Greenland, the Northern Pacific and Atlantic Oceans, and the Himalayas. This enhanced acid deposition may strongly impact local ecosystems. There is less sulfate deposition in East Asia, Southern Asia, India, Europe, North America and Central Africa in SAG relative to the control as a result of reduced anthropogenic sulfate emission.

Compared with the same period of RCP8.5, SAG increases sulfate aerosol concentration, as well as its deposition (Figure 6f). The total deposition change between SAG and RCP8.5 shows similar patterns as in previous studies using the Goddard Institute for Space Studies ModelE (Kravitz et al., 2009) and University of L'Aquila Composition-Chemistry Model (Visioni et al., 2018), even though in both previous cases equatorial $SO_2$ injection was simulated. The strongest deposition increase is over the North Atlantic, the mid Pacific and East Asia. This happens because, as shown by both previous papers, most of the changes in large-scale sulfate deposition would be driven by stratospheric circulation, resulting in most of the geoengineering particles crossing the tropopause at mid-latitudes and then being quickly removed via wet scavenging and gravitational settling, with little latitudinal mixing in the troposphere.

***Secondary Organic Aerosols***

SOA formation in WACCM is the result of prescribed gas-phase condensable emissions of SOA, so called SOAG, that are emitted at the surface and are directly proportional to the emissions of their precursor emissions (Liu et al., 2012). The emissions of SOAG are prescribed in the RCP8.5 pathway and the formation of SOA is therefore not coupled to climate-induced changes in biogenic emissions. Further SOAG are not assumed to undergo deposition in this model version. With the prescribed SOAG, MAM3 calculates condensation of SOAG as well as evaporation to output aerosol phase SOA in modes. The aerosol phase of SOA depends on (1) the total SOAG – when there is more SOAG, more condensation will produce more aerosol phase SOA; and (2) the equilibrium partial pressure of SOAG – when the equilibrium partial pressure is high, more SOAG would stay in gas phase and therefore less production of aerosol phase SOA. Temperature is positively correlated with the equilibrium partial pressure (Liu et al., 2012). Therefore, higher temperature will result in higher equilibrium partial pressure and tend to generate less aerosol phase SOA when other conditions are fixed.

Prescribed SOAG emissions increase until about 2050 and decline thereafter in both experiments (RCP8.5 and SAG). Under RCP8.5, the increasing of SOAG and temperature in the first couple decades are competing with each other to produce SOA. The global average aerosol phase SOA mass concentration shows increases from 2010 to 2030 in the Aitken mode and no change in the accumulation mode (Figures 7d and 7e). During this period, the increasing of SOAG balances out the negative impact from warming environment – therefore more SOAG produces more SOA. After 2030, SOA constantly decreases in both modes as a result of intensive warming and a much slower increasing rate of SOAG, which turns to a reduction around 2050. At the end of the 21$^{st}$ century, global averaged temperature under RCP8.5 is ~4.5°C warmer relative to the present. This warming effect in RCP85 tends to increase the equilibrium partial pressure of SOAG, and hence produces less SOA. Global averaged aerosol phase SOA reduces by 19% compared with the control (Figures 4 and 7).

On the other hand, under SAG, SOA shows a much weaker decline compared with RCP8.5. Since the two scenarios share the same prescribed SOAG emissions, the difference is only from temperature impacts on SOA formation. With geoengineering, temperature is not changing much in the troposphere relative to present. Therefore, SOA formation is mainly controlled by the SOAG amount and the decline is much smaller compared with RCP8.5. Deposition changes in total have mild impacts on SOA. Wet deposition of SOA in SAG is 1.1 Tg $a^{-1}$ less than that in RCP8.5 due to precipitation reduction, while dry deposition of SOA under SAG is 1.1 Tg $a^{-1}$ stronger than under RCP8.5 (Figure 4). The total removal of SOA is similar in RCP8.5 and SAG during 2080-2089. Therefore, changes in temperature in RCP8.5 result in less formation of SOA and hence a stronger decline in mass mixing ratio of SOA than in SAG.

***Black Carbon and Primary Organic Matter***

BC and POM emissions are prescribed in the model, including anthropogenic emissions (agricultural waste burning, domestic, energy, industry, ship, transportation, and waste treatment) and biomass-burning emissions (forest fire and grass fires) (Bond et al., 2007; Junker and Liousse, 2008). Anthropogenic emissions of BC and POM are strongly reduced under RCP8.5 but enhanced natural wildfires under global warming results in more carbon-related aerosol emissions. In total, future BC and POM emissions are less than the current level.

Climate changes of SAG have negligible impacts on the removal process of BC and POM. Hence their concentrations in SAG and RCP8.5 are close to each other with slightly lower values when geoengineering is applied (Figures 7f and 7g). However, Figure 4 shows that their wet and dry depositions are slightly less in SAG than in RCP8.5, which should result in a higher concentration of BC and POM under the geoengineering scenario. One possible explanation may be that the in-cloud wet and dry depositions of BC and POM are changing, but those are not saved in this model simulation. Values shown in Figure 4 for BC and POM are only wet and dry depositions below the cloud (Table 1). In the model, after emission BC and POM are instantaneously mixed with sulfate and other components in the accumulation mode, and thus have a higher wet removal rate by precipitation due to the high hygroscopicity of sulfate. Therefore, the unclosed budget of BC and POM in Figure 4 could be potentially completed by the overestimated in-cloud wet deposition.

***Dust***

Dust surface mass concentration is much lower under SAG than under RCP8.5 after year 2050 (Figures 7h and 7i) because of less dust emission (Figure 8a) and higher deposition rates (Figure 4) when geoengineering is applied.

In this study, the dust model used is based on the Dust Entrainment and Deposition Model (DEAD) (Zender et al., 2003). Dust mobilization depends on the wind friction speed and surface soil moisture, and its emission is highly determined by vegetation (Zender et al., 2003; Mahowald et al., 2006). Vegetation constrains dust emissions by reducing the wind drag on the erodible component of the surface and by increasing the soil moisture. The threshold for leaf area index (LAI) for the generation of dust is 0.3 $m^2/m^2$ (Mahowald et al., 1999; Okin, 2008). When LAI is equal to or larger than 0.3, dust emission will be terminated.

Global averaged dust emission decreases with time in RCP8.5 (Figure 8a) as a result of vegetation growth (Figures 8b and 8c), partly due to the $CO_2$ fertilization effect. Dust emission is further reduced under SAG (Figure 8a). Geoengineering does not change the $CO_2$ concentration in the atmosphere, but the cooler environment relieves plants from the heat stress in some regions. Meanwhile, although SAG results in less land averaged precipitation (Figure 8d), it reduces evaporation and evapotranspiration as well (Figures 8e, 8f and 8g). Hence, together with less run-off over the land under SAG geoengineering, there is more soil water available for vegetation growth under SAG (Figure 8i) (Cheng et al., 2019, submitted). A strong reduction of soil moisture in RCP8.5 is found, mainly in high latitudes in this model simulation. CLM4.5 simulates with a strong soil moisture reduction under global warming in permafrost regions – soil liquid water increase is much less than the melting of soil ice. Over the desert in Australia, one of the most important dust emission sources, precipitation increases (Figure 5a) as well as cooling in SAG relative to RCP8.5 results in more soil moisture. Although the averaged exposed one-sided Leaf Area Index shows a reduction when geoengineering is applied compared with RCP8.5 (Figure 8c), dust emission "hot spots" play an important role to determine the dust mass concentration. Therefore, we focus on the arid regions where dust is emitted, such as central Australia. The number of grid cells with LAI less than 0.3 with sulfate geoengineering is much less than that in RCP8.5 (Figure 9), which indicates that SAG has significantly terminated many dust emission "hot spots."

There is less dust wet and dry removal in RCP8.5 and SAG compared with the control (Figure 4), mainly due to less dust emission in the future. Compared with RCP8.5 (2080-2089), sulfate geoengineering tends to have slightly more dust removal (mainly wet in-cloud deposition) even with less dust emission. This enhanced removal results from the precipitation pattern change in SAG. Although global averaged precipitation reduces under SAG geoengineering, there is more precipitation over Central Australia, the east Indian Ocean along the Equator, and the Mediterranean Sea, where the dust concentrations are highest. Therefore, increased regional precipitation results in more wet deposition of dust.

***Sea Salt***

Sea salt is one of the most important natural aerosols, particularly in clean remote regions. The emission of sea salt depends on wind speed (e.g., Penner et al., 2001) and sea surface temperature (e.g., Mårtensson et al., 2003; Geever et al., 2005). In CESM, sea salt emission of particles with a diameter less than 2.8 μm is parameterized in terms of 10 m wind speed and sea surface temperature (Mårtensson et al., 2003).

Under RCP8.5, sea salt emission increases (Figures 7k and 7l) along with global warming, which is consistent with previous studies (Penner et al., 2001; Jones et al., 2007; Korhonen et al., 2010). Warming sea water tends to generate higher number concentrations of large particles and lower number concentrations of small particles (Mårtensson et al., 2003), so that warming tends to produce a higher mass concentration of sea salt. In addition, $CO_2$ induced global warming causes an increase in the pole-to-Equator temperature gradient, which enhances the westerlies. Stronger 10 m winds result in more sea salt emission in the model.

GLENS aims to use sulfate geoengineering to maintain the 2020 temperature gradient between the Equator and the poles. Therefore, the 10 m wind speed over high latitude oceans strongly decreases when geoengineering is applied (Figure 10), which is consistent with the reduction of sea salt emission in SAG relative to RCP8.5. Together with the cooling effect, there is less sea salt emitted under SAG geoengineering than RCP8.5. Although wet and dry depositions of sea salt are reduced as well under SAG, the reduction of emission is much stronger, and hence the surface mass concentration of sea salt in all modes decreases relative to RCP8.5.

## IV. Discussion and Conclusions

Our study shows that with the same future anthropogenic aerosol emissions, at the end of the 21st Century, global PM2.5 tends to decrease 8.7% with sulfate geoengineering, dominated by the changes of dust and sea salt. When dust and sea salt are not considered, global PM2.5 would show an increase of 7.7% mainly due to the increase of secondary organic aerosols. The strongest change in PM2.5 with geoengineering would occur in Oceania, with 25.3% reduction of PM2.5 and 18.3% increase of PM2.5 without dust and sea salt (Figure 11). In other populated regions, PM2.5 shows similar change patterns – reduction when dust and sea salt are the dominant factors, and increases when other components have greater importance. For example, in Asia, the highest population region, reduction of PM2.5 under sulfate geoengineering is largest over Southeastern Asia with a value of 8.5%, resulting from reduced sea salt, and over Central Asia (5.5%) resulting from reduction in dust (Figures 2 and 11).

Particulate matter is well known to affect morbidity and mortality (e.g., Ebi and McGregor, 2008; Brook et al., 2010). Increasing of PM2.5 together with tropospheric ozone under fixed unchanged anthropogenic aerosol emission would contribute 40% of the 1.1% mortality increase per degree temperature increase over the baseline rate (Jacobson, 2008). The simulated reduction of PM2.5 mass mixing ratio as well as possible reduction of surface ozone (Xia et al., 2017) when sulfate geoengineering would be applied indicate less human health stress due to PM2.5 exposure. Eastham et al. (2018) used Monte-Carlo simulations to estimate PM impacts on human health under sulfate geoengineering, and concluded that the cooling effect would increase PM mortality as a result of enhanced gas-aerosol exchange of $HNO_3$ as well as VOCs. This is consistent with our PM2.5 mass concentration change without dust and sea salt, but sulfate geoengineering would also significantly influence natural PM emission, which would change the sign of the change of total PM2.5 concentration. However, the composition of PM2.5 also plays an important role on human health (Franklin et al., 2008). Although the total PM2.5 mass concentration decreases in most regions under SAG, the composition change may counter the reduction of total mass concentration. Under SAG, percentages of sulfate, SOA, POM and BC in PM2.5 would increase (Figure 3). Franklin et al. (2008) showed that sulfate in PM2.5, along with temperature, explains 33% of PM2.5 mortality effect in 25 communities in the U.S. Brook et al. (2010) summarized that organic compounds such as SOA in PM2.5 are one of the components likely relevant to promoting cardiovascular disease. Therefore, sulfate geoengineering impacts not only the total mass concentration of PM2.5, but also its composition, which are both important to further understand its influence on human health.

A surprising result from this study is that although massive stratospheric injections of sulfur would be used in this scenario, it falls out mostly due to wet deposition (Figure 4), and does not have a substantial impact on surface PM2.5. While this enhanced acid deposition could certainly have consequences for natural ecosystems and human structures (not addressed here), the additional aerosol that makes its way into the boundary layer is in the coarse mode, and has very small increases, with negligible impact on PM2.5.

***Uncertainties***

All of the above results, while analyzed with a state-of-the-art climate model, depend on the model assumptions. Here we address those and their impacts on the results, and point the way toward future studies to address the effects of these assumptions.

For example, the model set-up restrains several important sources and processes that influence PM2.5 mass concentration under sulfate geoengineering. The biogeochemical emission model - the Model of Emissions of Gases and Aerosols from Nature (MEGAN) (Guenther et al., 1995; Guenther et al., 2012) only interacts with the atmosphere via CO, and all other VOCs are prescribed. Five of those prescribed VOCs are used to calculate gas phase SOA. Since isoprene and terpenoids are the most abundant VOC compounds (Guenther et al., 1995) and have a large biogenic source that is highly impacted by climate changes, the assumption here that emissions are fixed can introduce large biases in our results. Biogenic VOCs (BVOCs) are sensitive to temperature change. Under RCP8.5, BVOCs tends to increase with a warmer environment (Guenther et al., 2006; Arneth et al., 2010), while with the cooling from sulfate geoengineering, we expect reduced BVOC emission. Therefore, using prescribed VOCs, our results for SOA are underestimated in RCP8.5 and overestimated in SAG. The strong increase of SOA when sulfate geoengineering is applied might be much less, or even have an opposite sign.

Prescribed BC and POM exclude the climate change impacts on wildfire. Wildfires have a significant influence on PM concentration (Jacob and Winner, 2009) by emitting organic carbon and black carbon. In a warmer future, wildfire would contribute 36% and 16% more organic carbon and black carbon respectively to PM over the western U.S. during summertime (Spracklen et al., 2009). There has been no study on how sulfate geoengineering would impact wildfires yet. But one model study simulating a regional nuclear war between India and Pakistan showed that a drier and cooler environment may promote wildfires in the Amazon (Mills et al., 2014). In this study, prescribed BC and POM result in no change in BC and POM mass concentration in RCP8.5 and SAG geoengineering. If an interactive wildfire model were used, the different emissions would induce different BC and POM mass concentrations.

One conclusion of this study is the reduction of global averaged PM2.5 in SAG compared with RCP8.5, which is a result of reduced dust emission in SAG. Since dust emission is correlated with soil moisture and vegetation growth, this conclusion may highly depend on how well the land model that is used represents soil water.

This study also does not include a comprehensive tropospheric chemistry model. The model used only has simple gas phase $SO_2$ chemistry in the troposphere, which ignores chemistry of other components, such as isoprene, and chemistry in the aerosol phase. In addition,

MAM3 used in this long-term climate simulation has a simplified treatment of nitrogen components, which are not involved in atmospheric chemistry and physical processes. For example, Horowitz et al. (2007) showed that isoprene nitrate chemistry is highly uncertain and different model parameterizations would significantly change the productions from isoprene nitrate chemistry (Wu et al., 2008) and therefore impact on gas phase SOA formation.

Despite all uncertainties and imperfections, this is the first study to address PM2.5 concentration changes under sulfate aerosol geoengineering. More research is needed to investigate this topic comprehensively.

**Acknowledgements:**

This work is supported by NSF grant AGS-1617844. Computing and data storage resources were provided by the Computational and Information System Laboratory (CISL) at NCAR. We thank Yangyang Xu for helping with the PM2.5 calculation. And we thank Natalie M. Mahowald for helping us understanding dust emission in CESM. The GLENS project is supported by the Defense Advanced Research Projects Agency (DARPA) and all output are available to the community via the Earth System Grid (see information at http://www.cesm.ucar.edu/projects/community-projects/GLENS/).

**Table 1.** Model output of aerosol component in different sizes.

| Mode* | Components | DD[1] | WD[2] | DDC[3] | WDC[4] |
| --- | --- | --- | --- | --- | --- |
| Aitken 0.015-0.053 (μm) | $SO_4$ | x | | x | x |
| | SOA | x | x | x | x |
| | Sea Salt | | | x | x |
| Accumulation 0.058-0.27 (μm) | $SO_4$ | | x | x | x |
| | SOA | x | x | x | x |
| | BC | x | x | | |
| | POM | x | x | | |
| | Dust | | | x | x |
| | Sea Salt | | | x | x |
| Coarse 0.8-3.65 (μm) | $SO_4$ | | | x | x |
| | Dust | | | x | x |
| | Sea Salt | | | x | x |

* dry diameter size ranges from $10^{th}$ to $90^{th}$ percentiles (Liu et al., 2012)
[1]DD: Dry Deposition below cloud
[2]WD: Wet Deposition below cloud
[3]DDC: Dry Deposition in cloud
[4]WDC: Wet Deposition in cloud

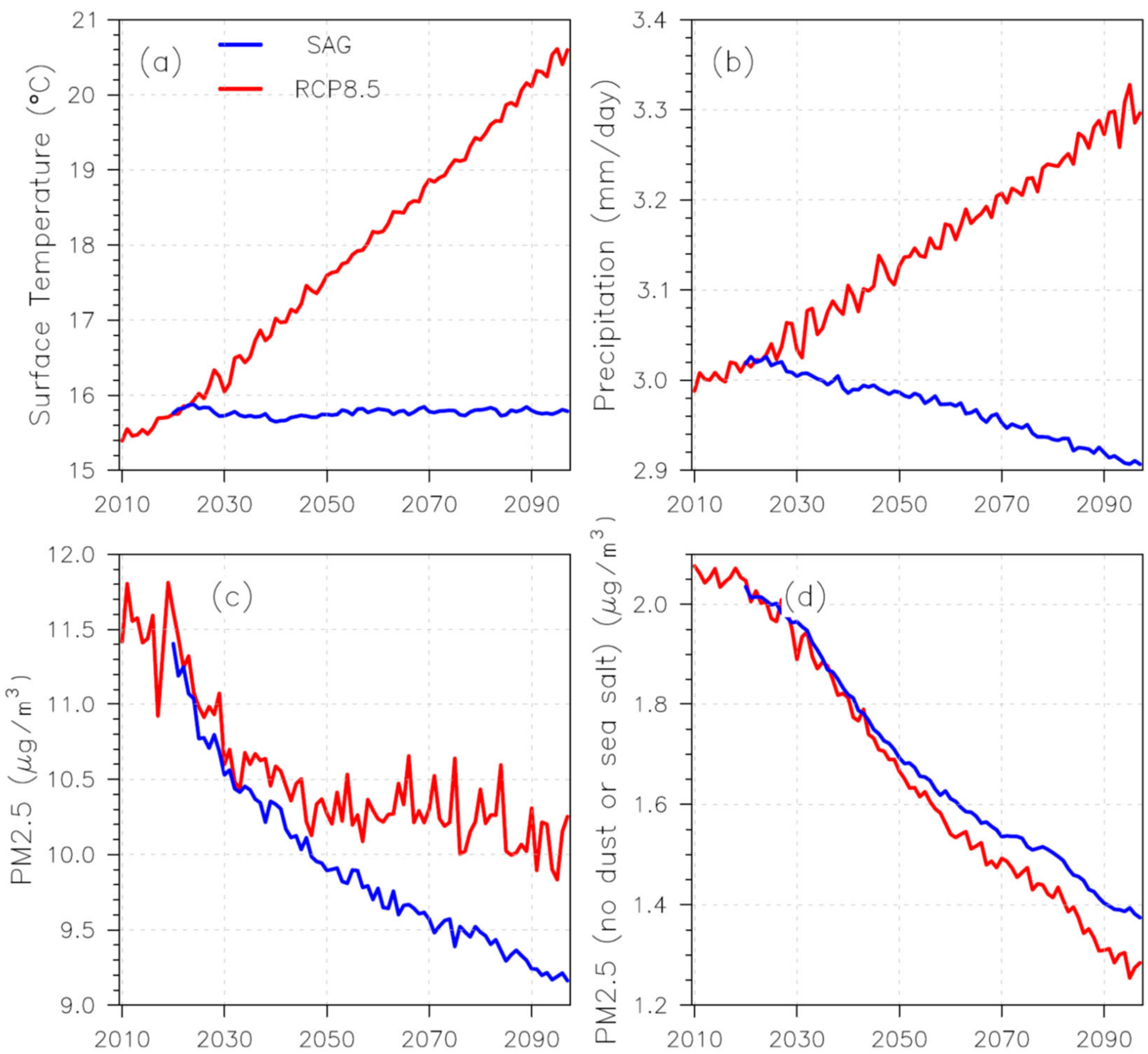

**Figure 1:** Global averaged (a) temperature (b) precipitation (c) PM2.5 and (d) PM2.5 without dust or sea salt. Lines are ensemble average: there are 20 ensemble members of RCP8.5 from 2010 to 2030, 3 ensemble members of RCP8.5 from 2031 to 2097, and there are 20 ensemble members of GLENS from 2020 to 2097.

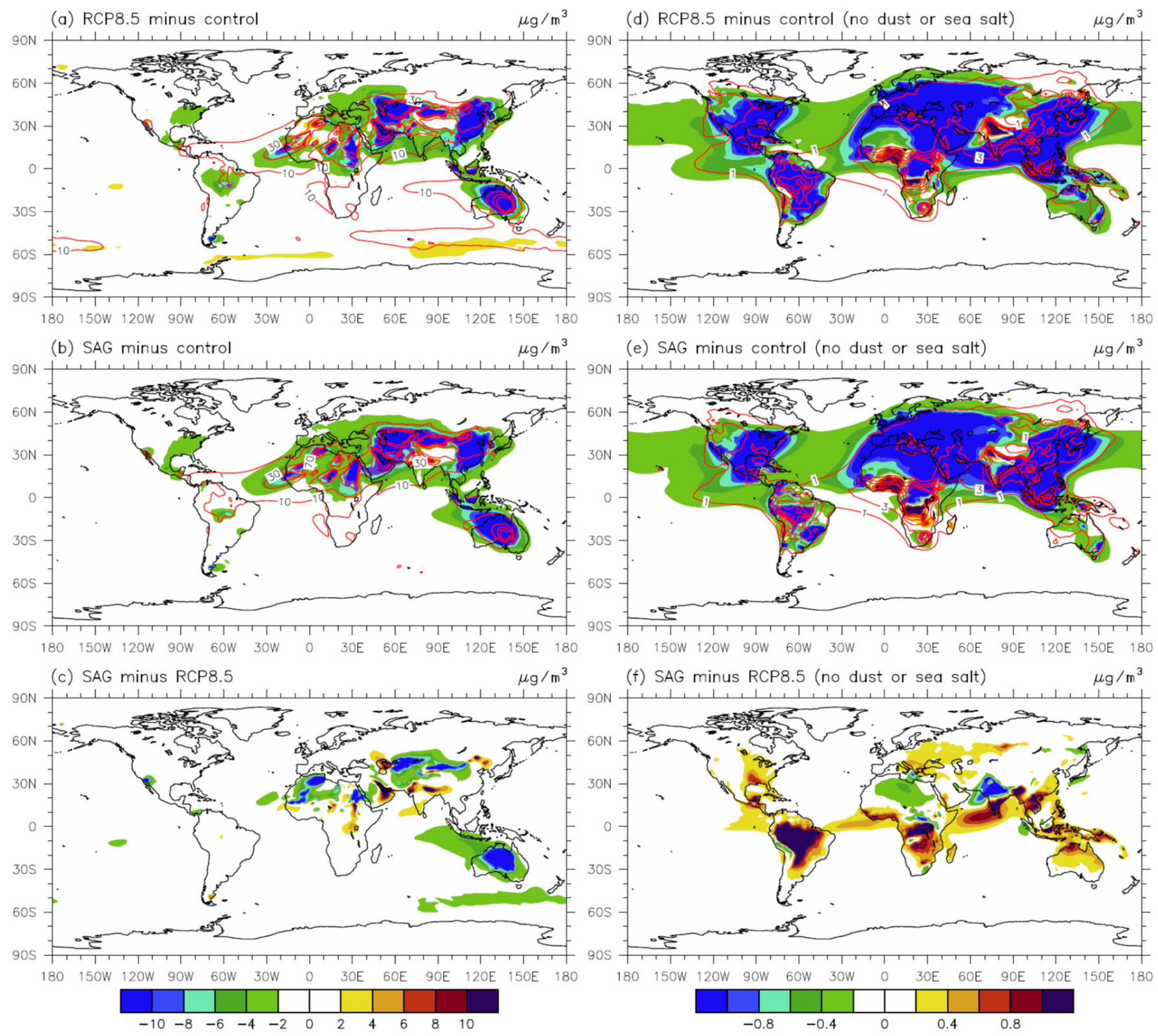

**Figure 2.** Differences of surface PM2.5 concentrations (with and without dust or sea salt) between RCP 8.5 (2080-2089) and RCP8.5 (2010-2019) (a and d), SAG geoengineering (2080-2089) and RCP8.5 (2010-2019) (b and e), and SAG geoengineering (2080-2089) and RCP8.5 (2080-2089) (c and f). The left panel is PM2.5, and the right panel is PM2.5 without dust and sea salt. The color bar on the left is 10 times than the one on the right. The red contour lines are the absolute values of PM2.5 concentration of (a) RCP8.5 (2080-2089) and (b) SAG (2080-2089), and the absolute values of PM2.5 concentrations without dust or sea salt of (d) RCP8.5 (2080-2089) and (e) SAG (2080-2089).

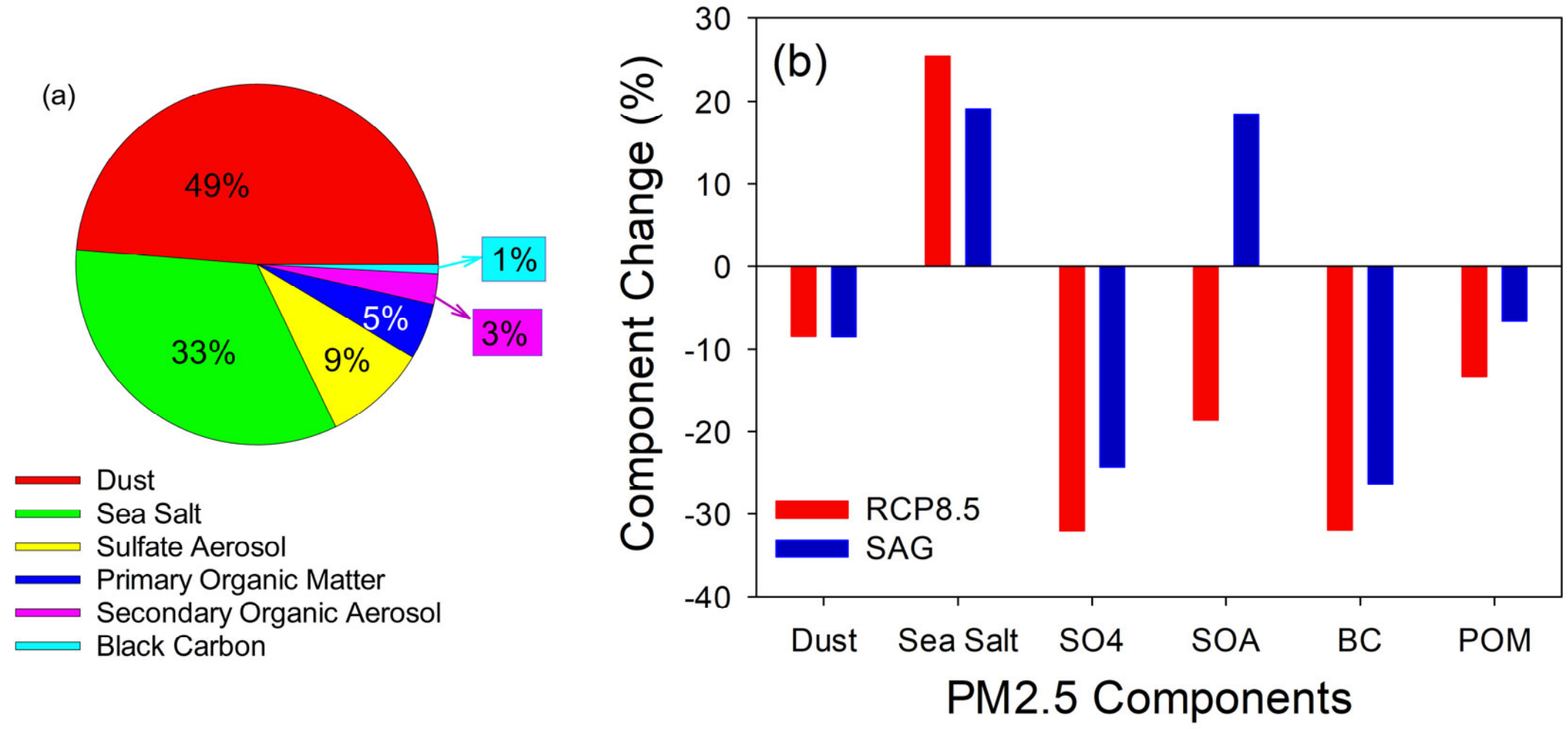

**Figure 3.** (a) Percentage of PM2.5 components during 2010-2019 for RCP8.5 and (b) percentage changes of PM2.5 components for RCP8.5 (2080-2089) and SAG (2080-2089) compared with RCP8.5 (2010-2019).

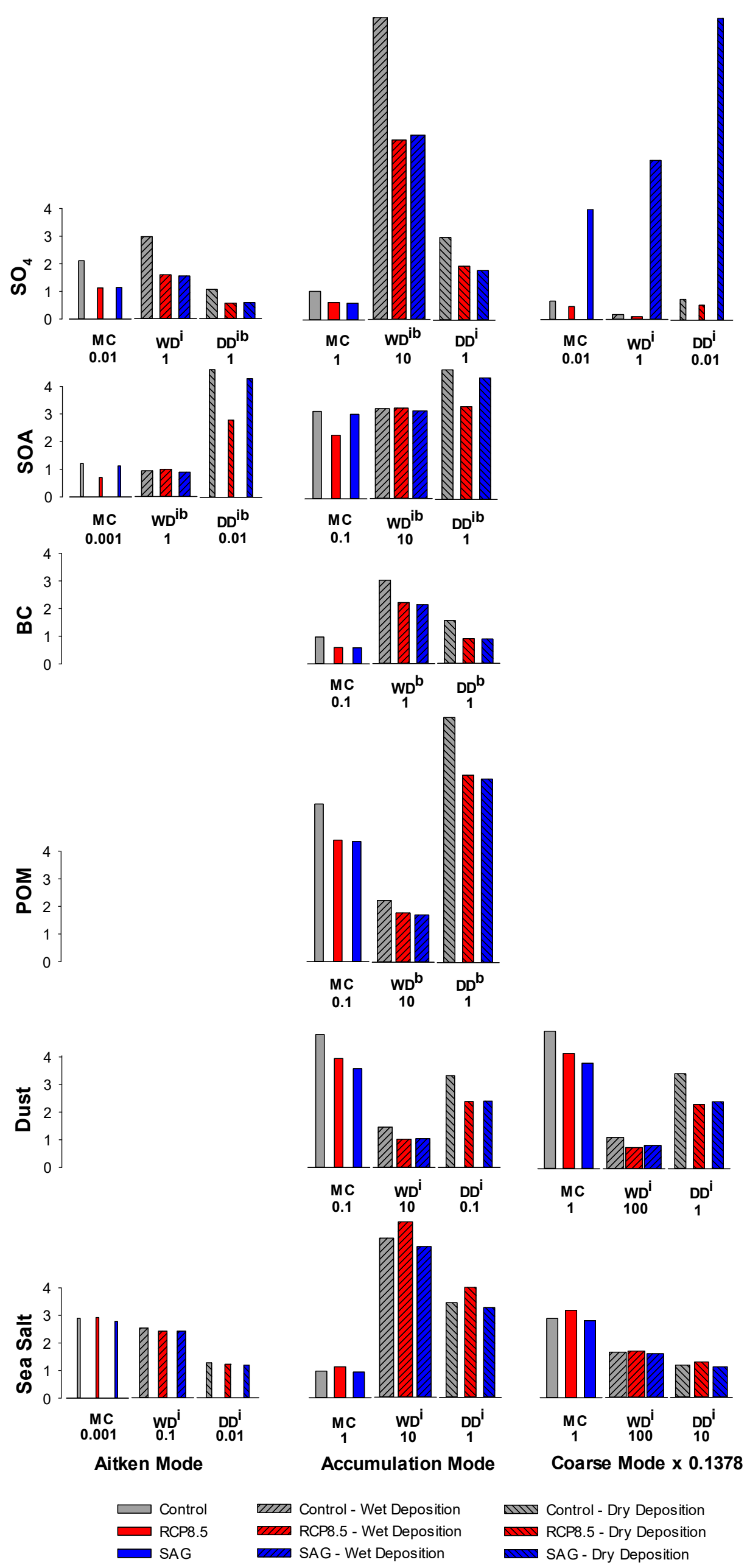

**Figure 4**. Global averaged aerosol surface mass concentration (MC), wet deposition (WD) and dry deposition (DD) in three modes. In WD and DD, $^{i}$ indicates in-cloud deposition, $^{b}$ indicates below-cloud deposition, and $^{ib}$ indicates in-cloud and below-cloud depositions. The unit of MC is μg/m$^3$ and of WD and DD is Tg/yr. WD and DD are the sum of in-cloud deposition and below-cloud deposition, if the values are available (Table 1). Labels on the y-axis need to be multiplied by the factors under the labels on the x-axis to get the actual values of MC, WD and DD. The width of the bars indicates the magnitude of the factors – the wider the larger (applied to the values on the y-axis).

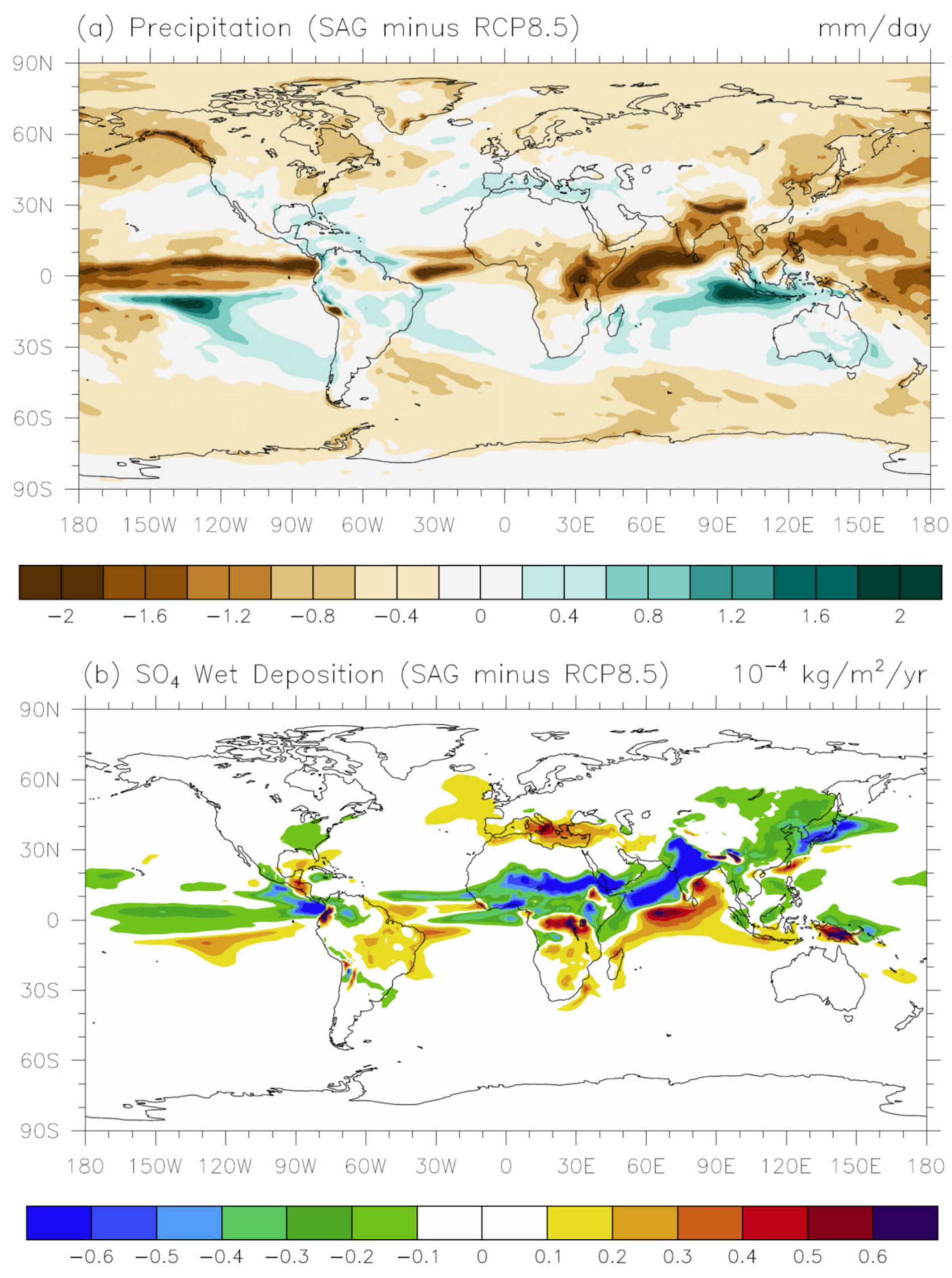

**Figure 5**. Differences of (a) precipitation and (b) surface sulfate below-cloud wet deposition in accumulation mode between SAG geoengineering (2080-2089) and RCP8.5 (2080-2089).

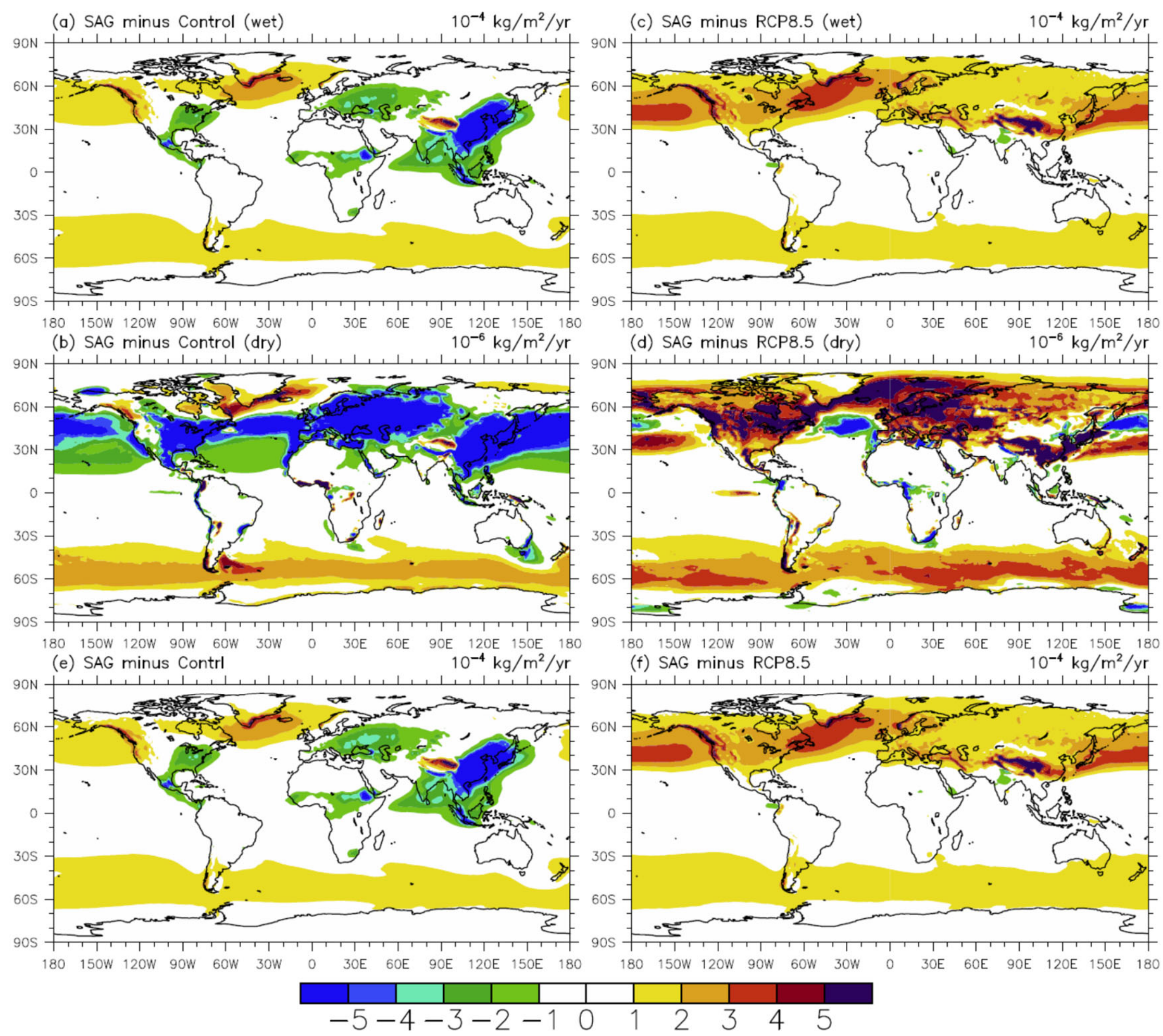

**Figure 6.** Sulfate aerosol wet, dry, and total deposition differences between SAG and control (a, c, e), and between SAG and RCP8.5 (b, d, f) for 2080-2089. Note that the unit of dry deposition (b and d) is 100 times less than wet deposition and the total deposition.

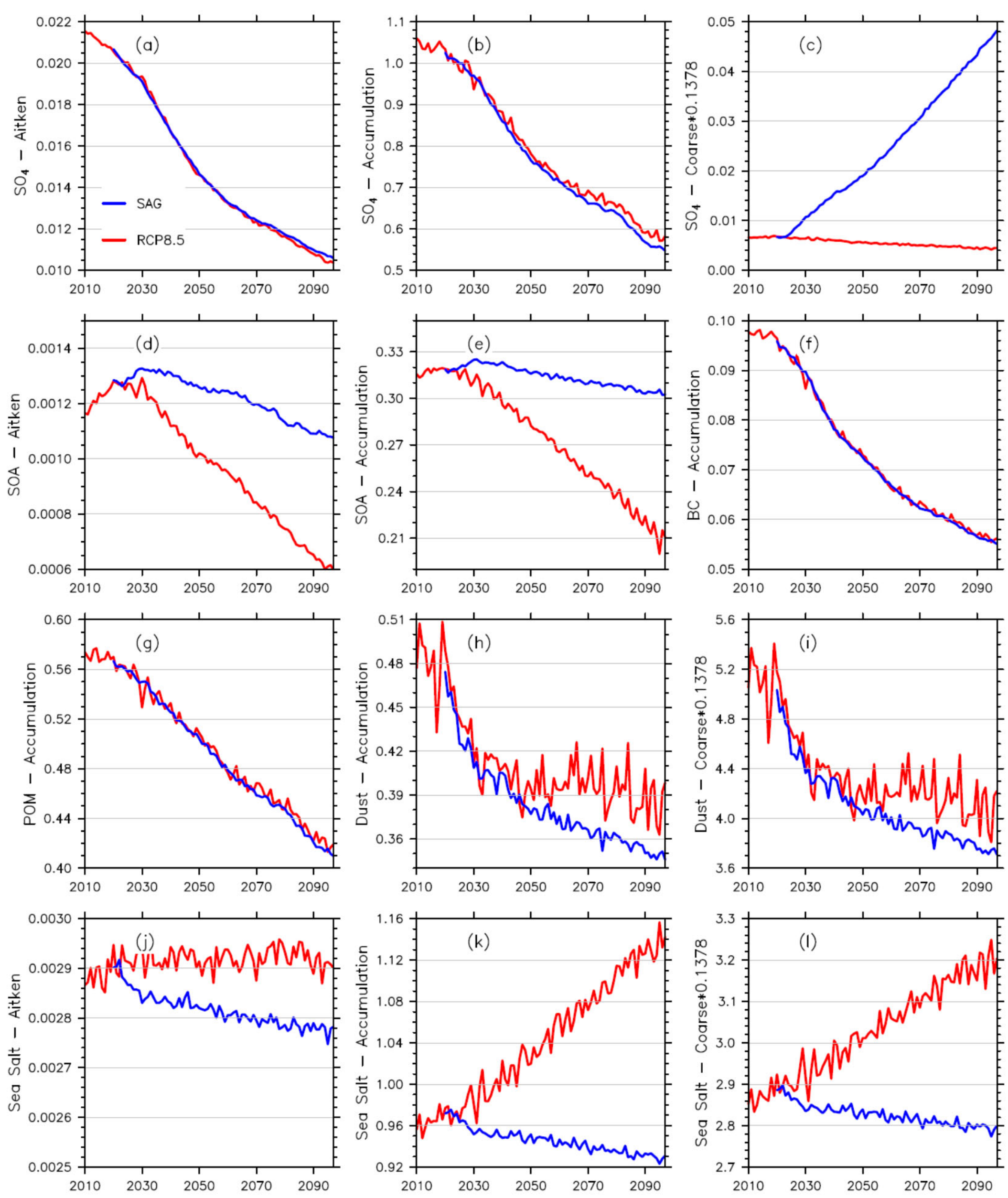

**Figure 7:** Global averaged surface mass concentrations (μg/m$^3$) of PM2.5 components in different modes: (a) sulfate aerosol in Aitken mode; (b) sulfate aerosol in accumulation mode; (c) sulfate aerosol in coarse mode; (d) SOA in Aitken model; (e) SOA in accumulation mode; (f) black carbon in accumulation mode; (g) primary organic matter in accumulation mode; (h) dust in accumulation mode; (i) dust in coarse mode; (j) sea salt in Aitken mode; (k) sea salt in accumulation mode; and (l) sea salt in coarse mode.

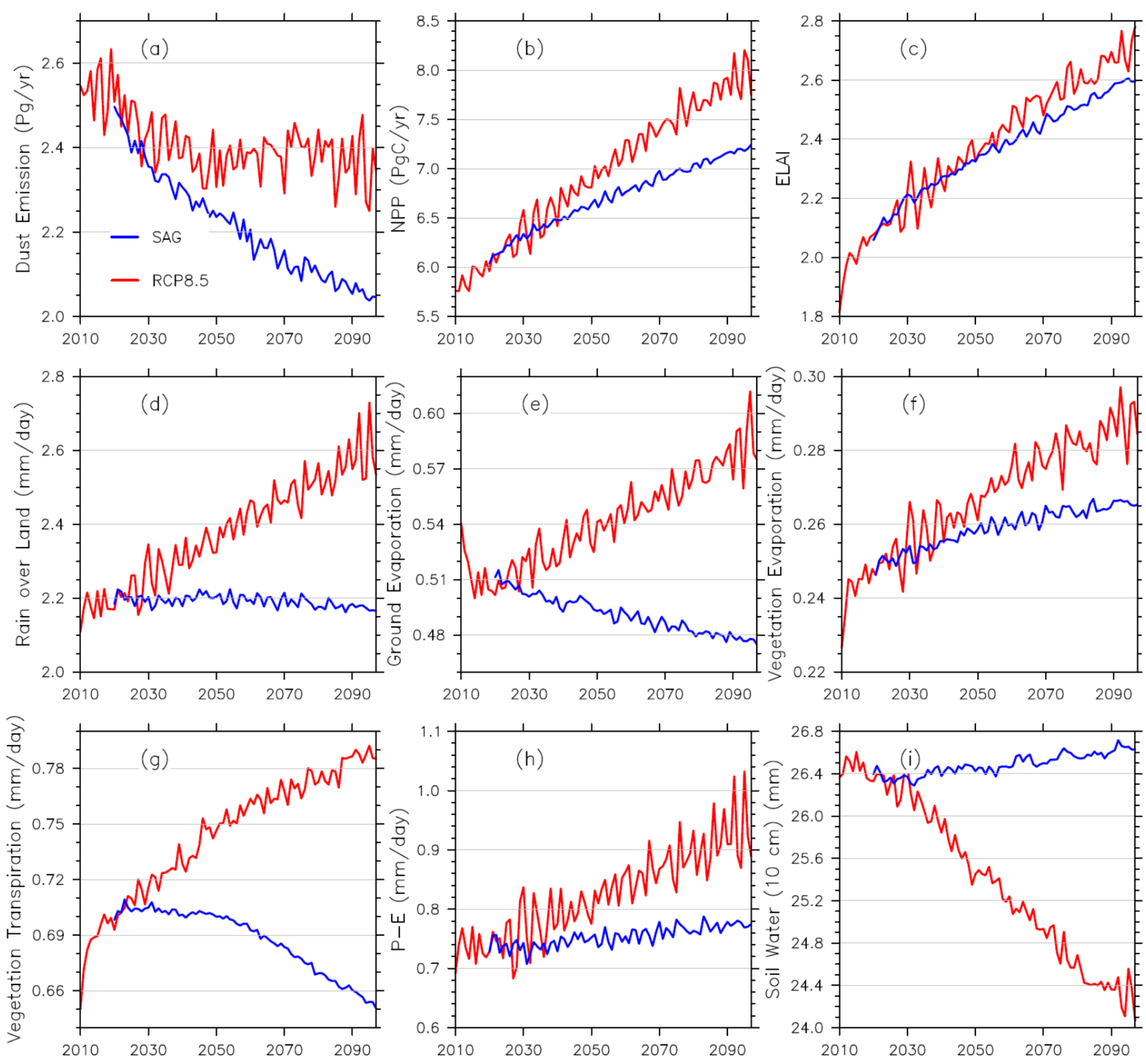

**Figure 8**: Land averaged (a) dust emission; (b) net primary productivity (NPP); (c) effective leaf area index (ELAI); (d) land precipitation; (e) ground evaporation; (f) canopy evaporation; (g) canopy transpiration; (h) land precipitation minus land evaporation and transpiration; and (i) soil water above 10 cm.

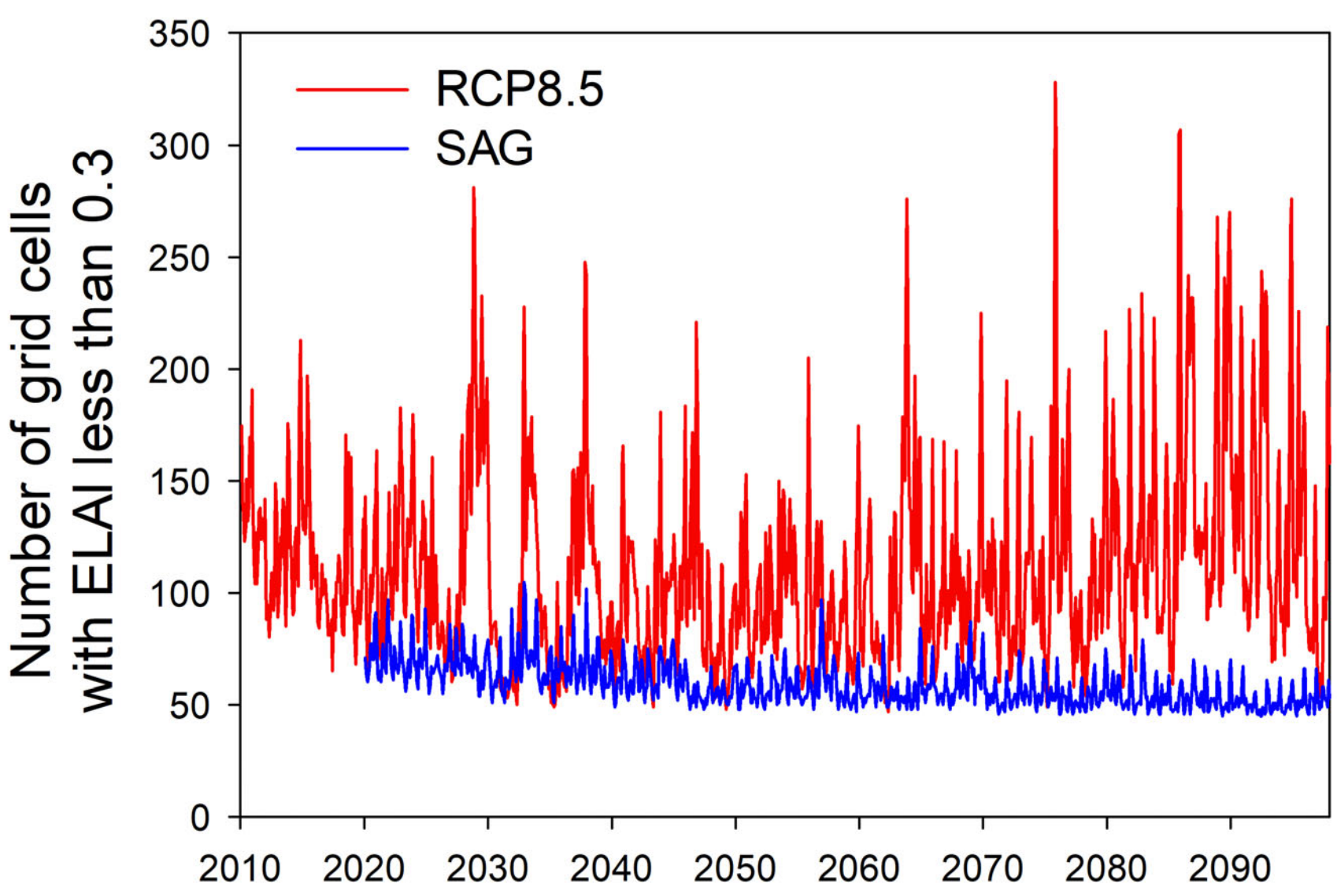

**Figure 9:** Number of grid cells with Exposed one sided Leaf Area Index (ELAI) less than 0.3 over Australia (15-35.5 °S, 115-152.5 °E) 0.3 is the threshold for vegetation friction to generate dust emission. When ELAI is larger than 0.3, dust emission is terminated.

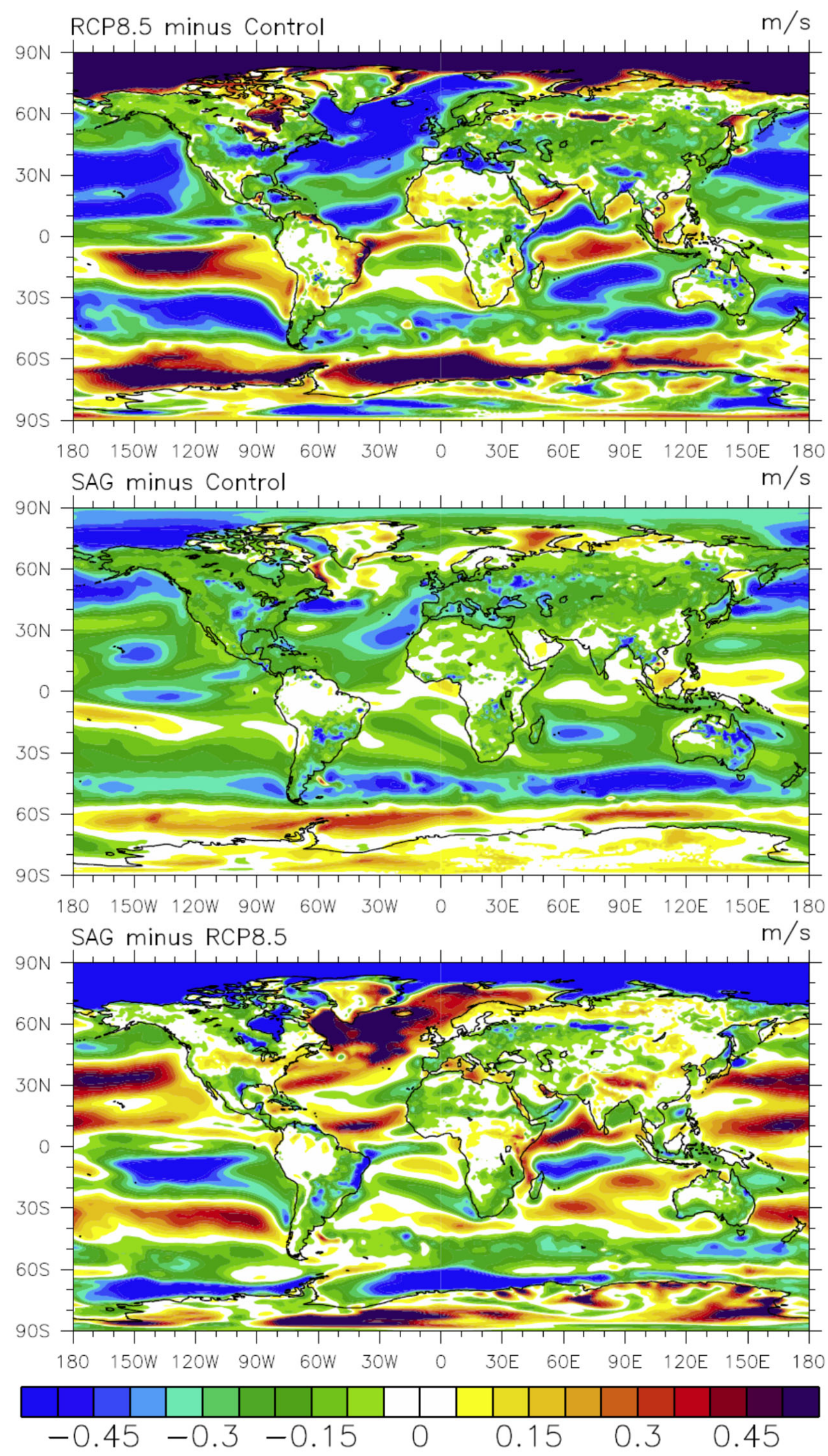

**Figure 10**: Differences of 10 m wind speed (m/s) (a) between RCP8.5 (2080-2089) and control; (b) between SAG (2080-2089) and control; and (c) between SAG (2080-2089) and RCP8.5 (2080-2089).

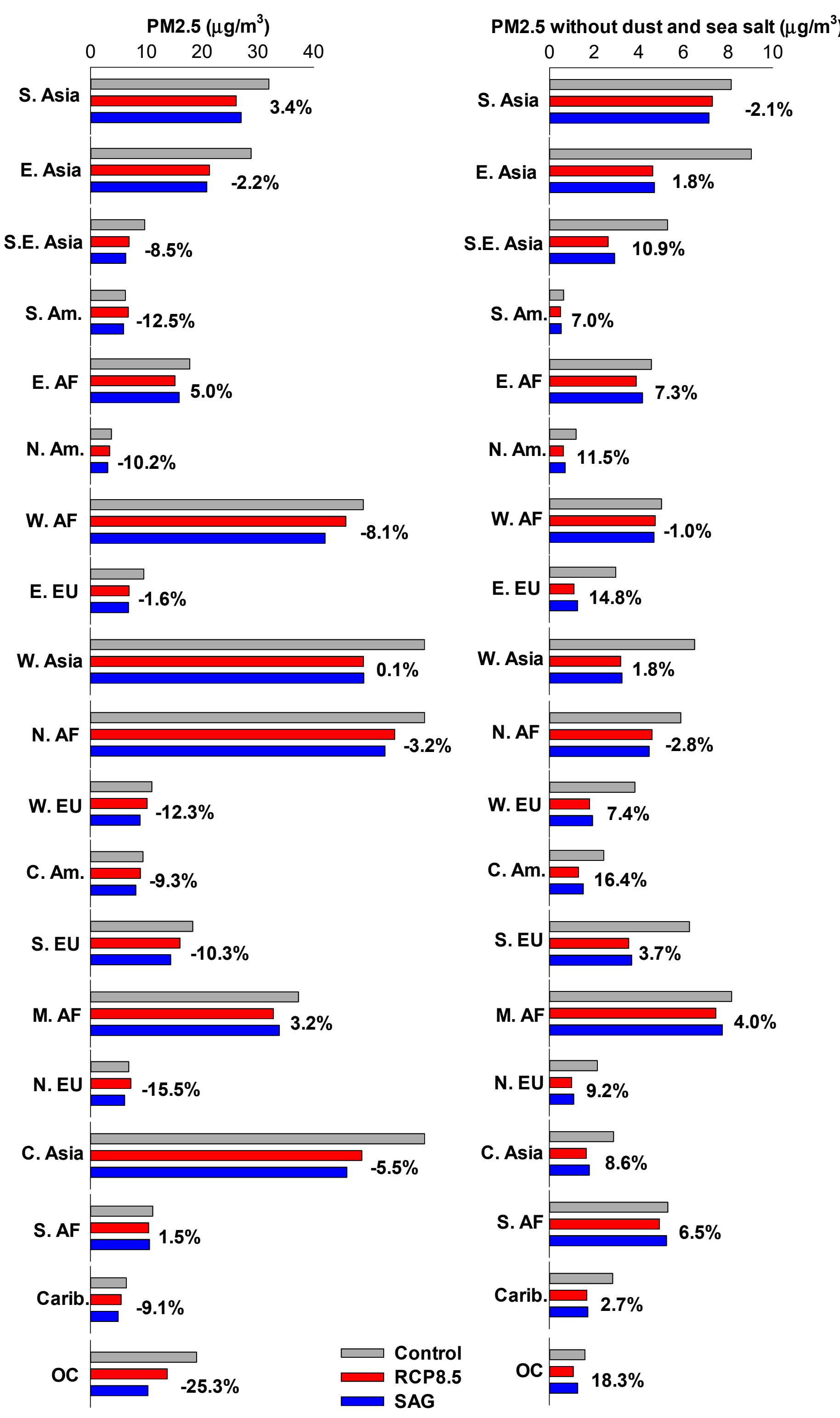

**Figure 11**: PM2.5 mass concentration in 19 regions in control for RCP8.5 (2010-2019) (gray), RCP8.5 (2080-2089) (red), and SAG geoengineering (2080-2089) (blue). The order of regions is based on population. S. Asia is Southern Asia, E. Asia is Eastern Asia, S. E. Asia is Southeastern

Asia, S. Am. is Southern America, E. AF. is Eastern Africa, N. Am. is Northern America, W. AF. is Western Africa, E. EU is Eastern Europe, W. Asia is Western Asia, N. AF. is Northern Africa, W. EU is Western Europe, C. Am. is Central America, S. EU is Southern Europe, M. AF. is Middle Africa, N. EU is Northern Europe, C. Asia is Central Asia, S. AF. is Southern Africa, Carib. is Caribeean, and OC is Oceania.